\documentclass[aps,prl,twocolumn,superscriptaddress,showpacs,floats,preprintnumbers,amsmath,amssymb]{revtex4-1}

\usepackage{graphicx}% Include figure files
\usepackage{dcolumn}% Align table columns on decimal point
\usepackage{bm}% bold math

\begin{document}

\preprint{published in Phys. Rev. B 86, 085444 (2012) \copyright American Physical Society}

\title{Electronic structure of a subnanometer wide bottom-up fabricated graphene nanoribbon: End states, band gap and dispersion}

\author{C. Bronner}
\affiliation{Fachbereich Physik, Freie Universit{\"a}t Berlin,
Arnimallee 14, 14195 Berlin, Germany}

\author{F. Leyssner}
\affiliation{Fachbereich Physik, Freie Universit{\"a}t Berlin,
Arnimallee 14, 14195 Berlin, Germany}

\author{S. Stremlau}
\affiliation{Fachbereich Physik, Freie Universit{\"a}t Berlin,
Arnimallee 14, 14195 Berlin, Germany}

\author{M. Utecht}
\affiliation{Universit{\"a}t Potsdam, Institut f{\"u}r Chemie, Theoretische Chemie, Karl-Liebknecht-Stra\ss{}e 24-25, 14476 Potsdam, Germany}

\author{P. Saalfrank}
%\affiliation{Universi{\"a}t Potsdam, Institut f{\"u}r Chemie, Theoretische Chemie, Karl-Liebknecht-Stra\ss{}e 24-25, 14476 Potsdam, Germany}

\author{T. Klamroth}
\affiliation{Universit{\"a}t Potsdam, Institut f{\"u}r Chemie, Theoretische Chemie, Karl-Liebknecht-Stra\ss{}e 24-25, 14476 Potsdam, Germany}

\author{P. Tegeder$^{*}$}
\affiliation{Fachbereich Physik, Freie Universit{\"a}t Berlin,
Arnimallee 14, 14195 Berlin, Germany}

\date{published in Phys. Rev. B on 23 August 2012}

\begin{abstract}
   Angle-resolved two-photon photoemission and high-resolution electron
  energy loss spectroscopy are employed to derive the electronic
  structure of a sub-nanometer atomically precise
  quasi-one-dimensional graphene nanoribbon (GNR) on Au(111). We
  resolved occupied and unoccupied electronic bands including their
  dispersion and determined the band gap, which possesses an
  unexpected large value of 5.1 eV. Supported by density functional theory (DFT)
  calculations for the idealized infinite polymer and finite size
  oligomers an unoccupied non-dispersive electronic state with an
  energetic position in the middle of the band gap of the GNR could be identified.
  This state resides at both ends of the ribbon (end state) and is
  only found in the finite sized systems, {\em i.e.}  the oligomers.
\end{abstract}


\maketitle

\section{Introduction}
Along with the great interest in graphene that has emerged in recent
years \cite{Geim2007}, other low-dimensional carbon-based systems
attracted attention, one of which are graphene nanoribbons (GNRs).
These flat, narrow structures are quasi-one-dimensional and have been
subject to numerous theoretical
\cite{{Nakada1996},{Barone2006},{Son2006a},{Chaves2011}} as well as
some experimental studies
\cite{{Han2007},{Li2008},{Cai2010},{Tao2011}} since (unlike graphene
itself) they exhibit a tunable band gap over a wide range. This is
important for their possible implementation in nanoscale devices such
as transistors or transparent electrodes whereas they maintain the
large carrier mobility found in graphene sheets
\cite{{Geim2007},{Li2008}}.  Besides the width, one of the parameters
influencing the band gap (and the band dispersion along the ribbon
axis) is the edge shape. Depending on the corresponding
crystallographic direction in graphene, the ribbon edge's character
varies from armchair to zigzag quasi-continuously \cite{Ezawa2006},
much like in carbon nanotubes.
The band gap of GNRs are inversely proportional to the
ribbon width \cite{{Barone2006},{Ezawa2006}} yielding gap sizes up to
several electron volts in the sub-nanometer regime.  Although for a
wide range of ribbon widths, this inverse relationship has been
experimentally observed \cite{{Han2007},{Li2008}} and complies with
theory, atomically small widths ($<$1~nm) and the corresponding gaps
of more than 1~eV have not been observed which is due to the
preparation methods employed. Another property of interest is the carrier mobility in GNRs which
manifests itself e.g. in the one-dimensional band structure of the
ribbons which in first approximation resembles the band structure of
graphene projected onto the respective crystallographic direction
\cite{Nakada1996}. However, the lateral confinement of the electrons
in the ribbon alter the dispersion and particularly open a gap at the
Dirac point. Theory has dealt in great detail with the dispersion in
GNRs whereas there have been no corresponding experiments published so
far.

Synthesizing suitable GNRs is challenging, particularly for the
established methods such as lithography \cite{Han2007} and unzipping
carbon nanotubes \cite{Kosynkin2009} since in both cases the widths
are usually larger than 10~nm and because of defects at the edges
influencing the electronic structure. A powerful alternative in the
fabrication of nanostructures is the on-surface synthesis
\cite{Grill2007}, which has been successfully applied for the
generation of atomically precise GNRs \cite{Cai2010}. Thereby the GNR
is formed \emph{via} a surface- and thermally assisted
\cite{Bjork2011} two-step process (see Figure \ref{fig:vibHREELS}(a)),
in which the adsorbed precursor molecule 10,10'-dibromo-9,9'-bianthryl
is dehalogenated followed by C--C coupling to form a polymer and finally
cyclodehydrogenation yields the defect-free armchair GNRs with a
well-defined width of 0.7 nm.

In this letter, we utilize angle-resolved two-photon photoemission
(2PPE) and high resolution electron energy loss spectroscopy (HREELS)
to determine the electronic band structure, i.e. unoccupied and
occupied electronic states as well as the band gap and dispersion for
the GNR shown in Figure \ref{fig:vibHREELS}(a) adsorbed on Au(111).
Supported by DFT calculations we demonstrate that the band gap of the
ribbon is surprisingly large namely 5.1 eV and contradicts all
calculated values known from literature
\cite{{Nakada1996},{Barone2006},{Son2006a},{Chaves2011}}. An
unoccupied electronic state is found in the band gap which possesses
no dispersion and originates from the molecular frontier orbitals
localized at the ends of the GNR with a finite length.

\section{Results and Discussion}
\begin{figure}[htb]
\centering
\resizebox{0.95\hsize}{!}{\includegraphics{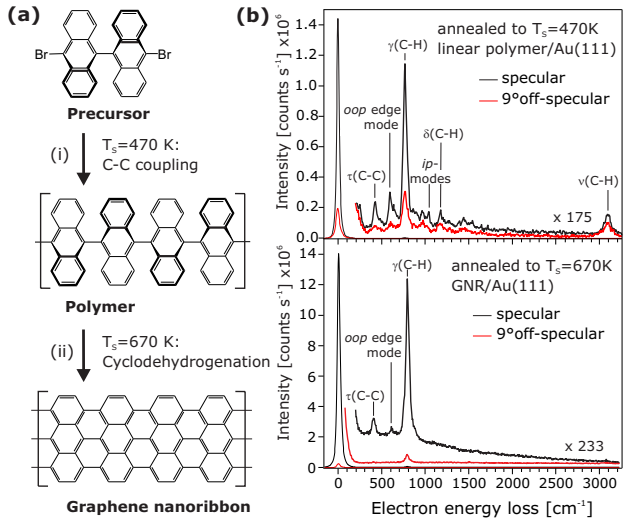}}
\caption{(a) Scheme of the surface-supported fabrication of graphene
  nanoribbons (GNRs) \emph{via} a two step process: (i) dehalogenation
  of precursor molecule 10,10'-dibromo-9,9'-bianthryl followed by a
  C--C-coupling and (ii) cyclodehydrogenation. (b) Changes in the
  vibrational HREEL spectrum observed during the cyclodehydrogenation
  step from the linear polymer to the nanoribbon measured with a
  primary electron energy of 3.5~eV.}
\label{fig:vibHREELS}
\end{figure}

The thermally activated surface-supported formation steps of the GNR
on Au(111) can nicely be followed by angle-resolved vibrational
HREELS, since well-defined and pronounced changes in the vibrational
spectrum are observed for each step. Fig. \ref{fig:vibHREELS}(b)
shows the data for both the polymeric phase and the aromatic
nanoribbons.
%
% Polymer phas
In the specular spectrum of the linear polymer, we observe several
out-of-plane molecular vibrations, namely the C--H wagging mode
($\gamma$(C--H)) at 758~cm$^{-1}$, an edge mode at 594~cm$^{-1}$ and
the phenyl ring torsion modes ($\tau$(C--C)) at 424~cm$^{-1}$. All
these modes show a drastic decrease in the spectrum measured in
off-specular geometry indicating that their intensities are
predominantly originating from dipole-scattering, i.e., they are
dipole active. Thus, the corresponding dipole moment changes during
vibration can be inferred to lie perpendicular to the surface. This
points towards a mainly flat adsorption geometry of the phenyl rings
in the polymeric phase. Due to the single C--C bonds connecting the building blocks of the
polymer and the steric
hinderance of the C--H groups, they are slightly tilted as observed in STM \cite{Cai2010}.
Indeed, at 3055~cm$^{-1}$ the C--H stretch mode ($\nu$(C--H)) shows a weak dipole activity. Accordingly this vibration exhibits a
component of the dipole moment change parallel to the surface normal
which is due to a tilting. Hence, our data fully supports the
adsorption geometry reported by Cai \emph{et al.} \cite{Cai2010}. Upon
cyclodehydrogenation and thus formation of the nanoribbons, the
vibrational spectrum is changed significantly.  The number of
molecular vibrations found in the spectrum is reduced to three modes.
The out-of-plane modes, the phenyl rings torsion mode $\tau$(C--C)
(404~cm$^{-1}$), the edge mode (609~cm$^{-1}$) and the bend mode
$\gamma$(C-H) at 793~cm$^{-1}$ are all purely dipole-active. This
clearly demonstrates that all phenyl rings are now orientated parallel
(flat-lying) to the surface as expected for the aromatic GNR.
By adsorption of xenon onto the so prepared GNR-covered surface, we may
roughly infer to the coverage of the nanoribbons on the gold surface: in
temperature programmed desorption, two desorption peaks are observed
which do not grow in intensity, as the layer thickness is increased to
very high coverages. These features are assigned to Xe desorbing from
both between and on top of the nanoribbons, respectively. From their
peak intensities we can roughly infer to a coverage of approximately 2/3 ML.

\begin{figure}[htbp]
\centering
\resizebox{0.7\hsize}{!}{\includegraphics{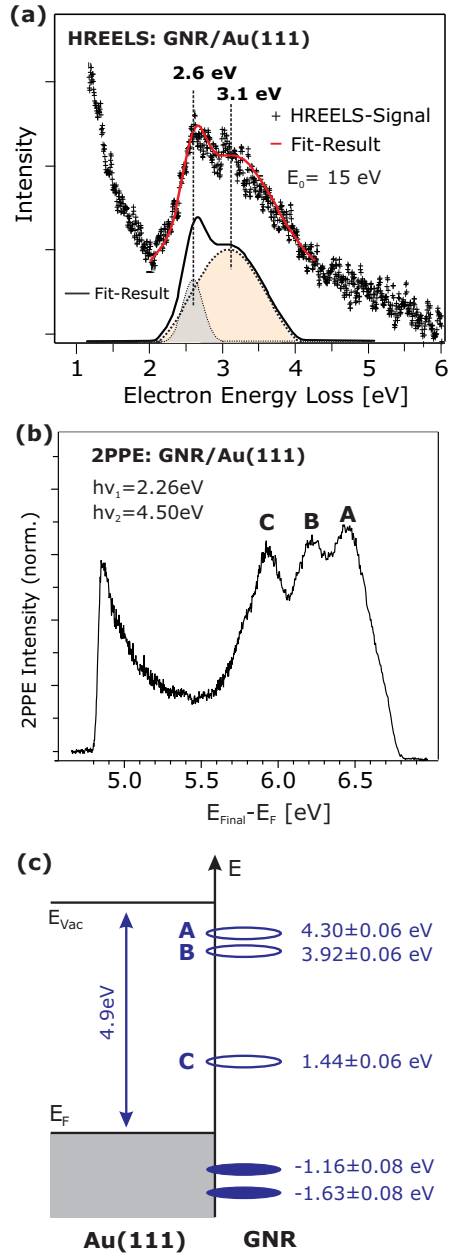}}
\caption{(a) Electronic HREEL spectrum of the GNR-covered gold surface
  recorded at a primary electron energy of 15~eV. Two electronic
  transitions are observed which are fitted using two Gaussian peaks.
  (b) Two-color 2PPE spectra recorded at photon energies of 2.25 and
  4.50 eV of GNR/Au(111). (c) Energetic position of GNR-derived
electronic states. The Fermi level of Au(111)
serves as reference. \label{eleStructure}}
\end{figure}

Electronic HREELS as well as 2PPE is employed to gain insights into
the occupied and unoccupied
GNR-derived electronic states and therefore the band gap. Figure \ref{eleStructure}(a) shows the result obtained from electronic
HREELS using a primary electron energy of $E_0$=15\text{eV}. A double
peak structure is observed which we fitted with two Gaussian peaks
yielding transition energies of 2.6~eV and 3.1~eV, respectively.
An exemplary two-color 2PPE spectrum recorded with 2.26 eV and 4.50 eV
photons is displayed in Figure \ref{eleStructure}(b). Several peaks are
observed and on the basis of photon energy dependent measurements
\cite{Hagen2010, Bronner2012} they can be related to photoemission from unoccupied
intermediate states. The peaks labeled with A and B are both populated
with the 4.50 eV photons and probed by $h\nu$=2.26 eV, thus they are
located at 4.30 $\pm$ 0.06 eV and 3.92 $\pm$ 0.06 eV, respectively,
with respect to the Fermi level ($E_{F}$). Note that the energetic positions given here are averaged values based on several measurements.
The peak labeled C
originates from a state which is probed by the 4.50 eV photons,
therefore it possesses an energetic position of 1.44 $\pm$ 0.06 eV
with respect to $E_{F}$.  Combining the HREELS transition energies
with the 2PPE results, we conclude that there are two occupied
electronic states lying at -1.16 $\pm$ 0.08 eV and -1.63 $\pm$ 0.08 eV
with respect to $E_{F}$. Since these energies do neither comply with
the bulk band structure of Au(111) nor with the Shockley surface
state, these occupied states can be assigned to the GNR. Figure \ref{eleStructure}(c) summarizes the the binding energies of all electronic
states observed in the present study.  On the basis
of the obtained data one would establish a band gap of 2.6 eV as found
as the lowest transition in the electronic HREELS measurement, which is in reasonable agreement with a recent measurement of the band gap on spatially aligned GNR
using ultraviolet and inverse photoemission \cite{Linden2012}.
However, for an unambiguous identification of the band gap one needs
further information on the nature of the unoccupied bands (states).
In order to gain deeper insights into the properties we performed
dispersion measurements using angle-resolved 2PPE, which provide
information about the extent of electron delocalization/localization.

\begin{figure}[htb]
\centering
\resizebox{0.95\hsize}{!}{\includegraphics{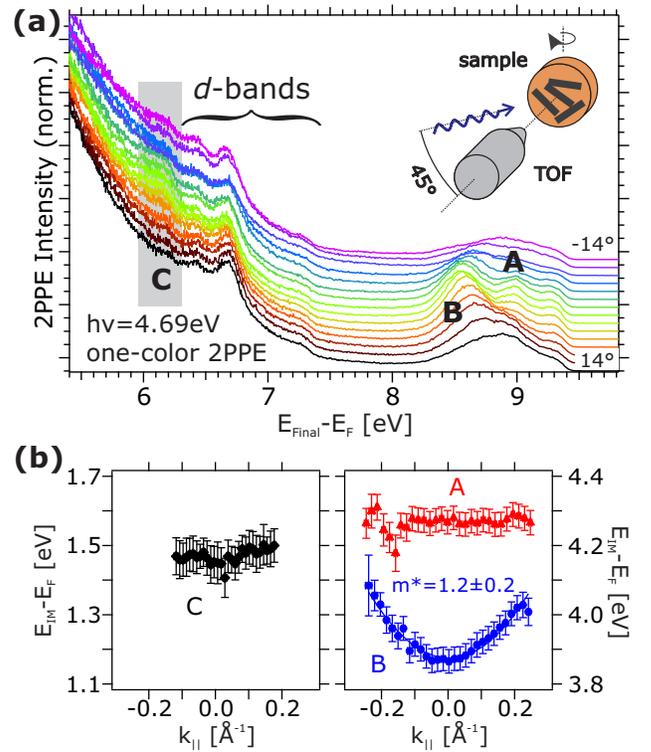}}
\caption{(a) Angle-resolved one-color 2PPE spectra (Inset: Measuring geometry in angle-resolved 2PPE experiments. Among the randomly oriented GNRs on the surface mainly those lying in the plane perpendicular to the rotation axis of the sample are detected for off-normal emission).  (b): Dispersion of the spectral feature labeled as A,  B and C. While the states A and C are localized, state B shows a strong dispersion, \emph{viz.} is delocalized.
  \label{fig:dispersion}}
\end{figure}
As observed in STM \cite{Cai2010}, the GNRs investigated in this study
are oriented randomly on the Au(111) surface. In our experiment, the
momentum information is obtained by rotating the sample in front of the time-of-flight
(TOF) spectrometer as seen in Fig. \ref{fig:dispersion}(a). While all
nanoribbons contribute to the signal detected in normal emission, this
is not the case as we rotate the sample.  A photoelectron emitted from
the one-dimensional band structure of GNR band has two contributions to its momentum  $k_{\|}$ parallel to
the surface, i.e. the component along the ribbon axis (which is the
quantity of interest) and a component perpendicular to it which is
random (as the electrons are localized in this direction).  Thanks to the
measuring geometry our TOF  only detects photoelectrons emitted in
the plane perpendicular to the rotation axis. This can either be
electrons emitted along the axis of ribbons lying in this plane or
electrons which (due to the random momentum perpendicular to the
ribbon axis) are emitted at an angle with respect to the ribbon which
is equal to the angle between this GNR and the detection plane. In the
latter case, the detected momentum will be higher than the momentum
the electron had in the band which causes the measured dispersion to
be smeared out to higher \emph{k} but not in energy.  Since for finite
angles only ribbons lying perpendicular to the rotation axis can be
measured, we expect a significant drop in photoemission intensity away from the $\Gamma$ point.

From the three unoccupied bands observed in 2PPE only the one labeled
as B shows a dispersion, \emph{viz.} the states A and C are localized.
Figure \ref{fig:dispersion}(a) displays a series of
one-color 2PPE spectra recorded at a photon energy of 4.66 eV for various
angles between the surface normal and the TOF axis. Beside the contributions of the gold \emph{d}-bands, the
peaks labeled A, B,  and C are seen. While B exhibits a strong dispersion
 around the $\Gamma$ point, state A and C show no dispersion (see
Fig. \ref{fig:dispersion}(b)). The parabolic behavior of state B complies qualitatively with all
theoretical calculations on armchair nanoribbons. The effective mass ($m^{*}$)
of $1.2 \text{ m}_\text{e}$ is significantly higher than theoretical
models generally predict for a variety of GNRs
\cite{{Wang2011},{Yang2007}}.

However most surprising the lowest unoccupied state located at 1.44 eV
above $E_{F}$ (peak C) exhibits no dispersion, therefore we assign
the first dispersive state (state B) above the Fermi level, which
possesses an energetic position of 3.92 eV to the conduction band of
the GNR. Thus, with the valence band located at -1.16 eV with respect
to $E_{F}$ the band gap is 5.1 eV. This value is significantly higher
than the calculated values lying in the range between 1.5~eV and
2.1~eV depending on the applied method \cite{Cai2010}. For the use in
technological applications, the size of the band gap plays a key role;
the one we observe for the studied armchair GNR is too large for instance
the use in a nanowire or even a transistor where a small gap is
preferred.

In order to understand the nature of the localized state C we
performed DFT calculations for the idealized infinite polymer and
finite size oligomers employing the PBE GGA-functional and the range
separated hybrid functional CAM-B3LYP, which is particularly suitable for long chain molecular systems (polyene) \cite{Yanai2004}
together with the 6-311G** basis
set as implemented in the Gaussian~09 program package \cite{g09}. All calculations
were done for the GNR alone, {\em i.e.}  without the supporting
Au(111) surface, and the geometry of the GNR was fully optimized. For
the periodic calculations a $k$-point grid with 240 points in the
first Brillouin zone (1BZ) and a real space cut off of 600 bohrs were used.
The PBE results obtained here compare very well with calculations
reported by Cai and coworkers \cite{Cai2010}. Furthermore, for the periodic
calculations, the geometry of the GNR depends only very weakly on the
choice of the functional. For instance, we obtain a lattice constant
of 4.310~\AA{} for PBE and for CAM-B3LYP we get 4.277~\AA{}. However, the band gap changes drastically due to the admixture
of exact exchange in the CAM-B3LYP functional. Using the PBE
functional results in a gap of 1.54 eV, while for CAM-B3LYP a gap of
4.35 eV is obtained. The same qualitative trend is observed when
using PBE0 \cite{Adamo1999}, i.e., PBE with an admixture of exact exchange, where we
obtain a band gab of 2.7 eV. Also, it has been shown, that
CAM-B3LYP is especially suitable for structural and electronic
properties of large unsaturated hydrocarbons \cite{Peach2007}, which is in line with the much
better agreement of CAM-B3LYP with the experimental results reported
here.

However, apart from the wider band gap and the larger dispersion, the
band structures for the PBE and CAM-B3LYP are qualitatively similar
and no indication of a non-dispersive state corresponding to the
localized state C could be found in the periodic calculations for the
idealized infinite polymer.
\begin{figure}[htb]
\centering
\resizebox{0.95\hsize}{!}{\includegraphics{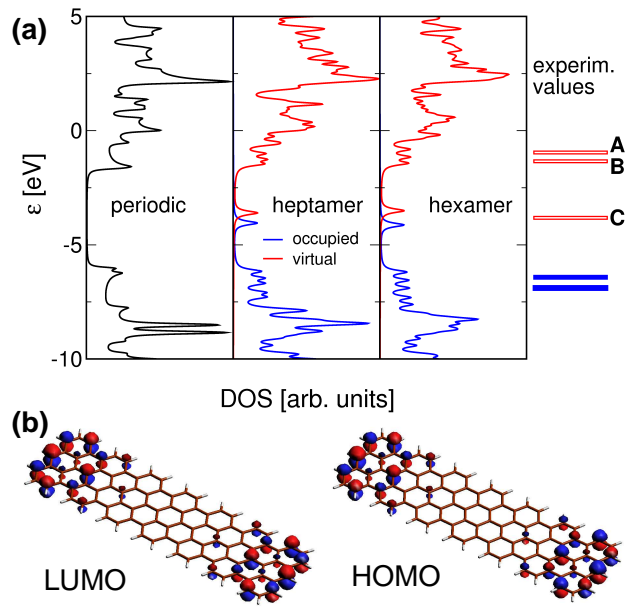}}
\caption{(a) Density of state obtained be Lorentzian-broadening for
  the periodic case (black line) and for the occupied (blue lines) and
  virtual orbitals (red lines) of the heptamer and the hexamer
  together with the experimental energies, which have been aligned to the middle of the band gap of the periodic GNR.  (b) HOMO and LUMO for the heptamer, which are localized at
  the ends of the oligomer (all calculations at the CAM-B3LYP/6-311G**
  level of theory).\label{fig:hvDep}}
\end{figure}
Therefore, we carried out additional calculations for oligomers up to
the heptamer, where the oligomers were saturated with hydrogens at
both ends and again a full geometry optimization was performed.  In
order to compare the electronic structure of the oligomers with the
one of the periodic structure, we computed the density of states (DOS)
by a simple Lorentzian-broadening procedure, where every Kohn-Sham
orbital energy or every crystal orbital energy within the 1BZ is
represented by a Lorentzian with a width $\Gamma$=0.2~eV for the
oligomers and $\Gamma$=0.1~eV for the periodic case.
Figure \ref{fig:hvDep}(a) shows the DOS for the periodic case (black
line) and for the occupied (blue lines) and virtual, {\em i.e.},
unoccupied, orbitals (red lines) of the heptamer and the hexamer using
the CAM-B3LYP functional.  As one can see, the DOS for the valence and
the conduction bands are
quite similar. The same holds for the geometric structure, as we get a
length of 4.278~\AA{} for the monomer unit in the middle of the
heptamer, which is nearly identical to the lattice constant of
4.277~\AA{} in the periodic case.

However, for the electronic structure of the oligomers we observe two
states inside the ``band gap'', namely the highest occupied (HOMO) and
lowest unoccupied molecular orbital (LUMO). These states are localized
at both ends of the oligomers, as shown in Fig. \ref{fig:hvDep}(b) for
the heptamer. Furthermore, the energy difference between these two
localized states decreases for longer chain lengths. In the experiments
GNRs with a length, which corresponds to
dozens of units (monomer length: 0.43 nm) \cite{Cai2010}, for which a non-resolvable energy difference
between the two localized states is to be expected. In addition, the
HOMO of the free GNR should also be unoccupied, when adsorbed at the
Au(111) surface, as its energy is above the Fermi level. Therefore,
these localized states can be identified as state C comparing the
calculated energies with experimental determined energies, which are
indicated at the right side of Fig. \ref{fig:hvDep}(a).

\section{Conclusion}

In conclusion, using two complementary surface-sensitive experimental
methods, namely HREELS and 2PPE, in combination with DFT calculations we were able to probe key electronic properties of a
defect-free graphene nanoribbon. We determined the band gap to be 5.1 eV, which is surprisingly high since previous calculations predicted much lower values.   A non-dispersive unoccupied electronic state of the GNR located in the band gap originates from both the HOMO and LUMO of the ribbons with a finite length (oligomer). These states are localized at both ends of the ribbon, thus we named them "end states". A controlled modification of the precursor molecule with suitable substituents or by doping may certainly open the perspective to create precise nanoribbons with lower band gaps, appropriate for applications.

\section{Acknowledgement}
We gratefully acknowledge financial support by the FU Berlin through the Focus Area "\emph{NanoScale}" and by the DFG through the collaborative research center SFB 658. We thank Cornelius Gahl (FU Berlin) for very fruitful discussions.

*petra.tegeder@physik.fu-berlin.de


\begin{thebibliography}{99}


\bibitem{Geim2007}
    {A. K. Geim and K. S. Novoselov, Nature Materials {\bf6}, 183-191 (2007)}
\bibitem{Nakada1996}
    {K. Nakada and M. Fujita, Phys. Rev. B {\bf54}, 17954-17961 (1996)}
\bibitem{Barone2006}
    {V. Barone, O. Hod, and G. E. Scuseria, Nano Lett. {\bf6}(12), 2748-2754 (2006)}
\bibitem{Son2006a}
    {Y.-W. Son, M. L. Cohen, and S. G. Louie, Phys. Rev. Lett. {\bf97}, 216803 (2006)}
\bibitem{Chaves2011}
    {A. J. Chaves, G. D. Lima, W. de Paula, C. E. Cordeiro, A. Delfino, T. Frederico, and O. Oliveira, Phys. Rev. B {\bf83}, 153405 (2011)}
\bibitem{Han2007}
    {M. Y. Han, B. \"Ozyilmaz, Y. Zhang, and P. Kim, Phys. Rev. Lett. {\bf98}, 206805 (2007)}
\bibitem{Li2008}
    {X. Li, X. Wang, L. Zhang, S. Lee, and H. Dai, Science {\bf319}, 1229-1232 (2008)}
\bibitem{Cai2010}
    {J. Cai \emph{et al.}, Nature (London) {\bf466}, 470-473 (2010)}
\bibitem{Tao2011}
    {C. Tao \emph{et al.}, Nature Physics {\bf7}, 616-620 (2011)}
\bibitem{Ezawa2006}
    {M. Ezawa, Phys. Rev. B {\bf73}, 045432 (2006)}
\bibitem{Kosynkin2009}
    {D. V. Kosynkin, A. L. Higginbotham, A. Sinitskii, J. R. Lomeda, A. Dimiev, B. K. Price, and J. M. Tour, Nature (London) {\bf458}, 872-876 (2009)}
\bibitem{Grill2007}
    {L. Grill, M. Dyer, L. Lafferentz, M. Persson, M. V. Peters, and S. Hecht, Nature Nanotechnology {\bf2}, 687-691 (2007)}
\bibitem{Bjork2011}
    {J. Bj\"ork, S. Stafstr\"omm, and F. Hanke, J. Am. Chem. Soc. \textbf{133}, 14884-14887 (2011)}

\bibitem{Hagen2010} S. Hagen, Y.   Lou, R.  Haag, M.  Wolf, P. Tegeder, New. J. Phys. \textbf{12}, 125022 (2010)
 \textbf{2010}, \emph{12}, 125022.

 \bibitem{Bronner2012}  Bronner, C.; Schulze, M.; Hagen, S.;  Tegeder P.
\emph{New J. Phys.} \textbf{2012},  \emph{14}, 043032.

\bibitem{Linden2012} S. Linden, D. Zhong, A. Timmer, N. Aghdassi, J.H. Franke, H. Zhang, X. Feng, K. M\"{u}llen, H. Fuchs, L. Chi, H. Zachariasm Phys. Rev. Lett. \textbf{108}, 216801 (2012).

\bibitem{Wang2011} G. Wang,  Phys. Chem. Chem. Phys., \textbf{13}, 11939-11945 (2011)


 \bibitem{Yang2007} L. Yang, C.-H. Park, Y.-W.  Son, M.L.  Cohen, S.G.  Louie, Phys. Rev. Lett., \textbf{99}, 186801 (2007)

\bibitem{Yanai2004} T. Yanai, D. Tew, N. Handy, Chem. Phys. Lett., \textbf{51}, 393 (2004)


\bibitem{g09}  Gaussian 09, Revision A.02, Frisch, M. J. \emph{et al.}, Gaussian, Inc., Wallingford CT, 2009.


\bibitem{Adamo1999} C. Adamo and V. Barone, J. Chem. Phys., \textbf{110}, 6158
(1999).

\bibitem{Peach2007} M.J.G. Peach \emph{et al.},
J. Phys. Chem. A, \textbf{111}, 11930 (2007).



\end{thebibliography}
\end{document}